\documentclass[runningheads]{llncs}
\usepackage{bbding}
\usepackage{amsmath,amssymb,amsfonts}
\usepackage{array}
\usepackage{booktabs}
\usepackage[colorlinks,linkcolor=blue,anchorcolor=blue,citecolor=blue]{hyperref}
\usepackage{graphicx}
\usepackage{epstopdf}
\begin{document}
\title{Interpretable Encrypted Searchable Neural Networks}
\author{Kai Chen\inst{1}\and
Zhongrui Lin\inst{2} \and
Jian Wan\inst{2}\and
Chungen Xu\inst{1(}\Envelope\inst{)}}
\authorrunning{Kai Chen et al.}
\institute{School of Science, Nanjing University of Science and Technology, Nanjing, CHN\\
\and School of Computer Science and Engineering, NJUST, Nanjing, CHN\\
\email{\{kaichen,zhongruilin,wanjian,xuchung\}@njust.edu.cn}}
\maketitle
\begin{abstract}
  In cloud security, traditional searchable encryption (SE) requires high computation and communication overhead for dynamic search and update. The clever combination of machine learning (ML) and SE may be a new way to solve this problem. This paper proposes interpretable encrypted searchable neural networks (IESNN) to explore probabilistic query, balanced index tree construction and automatic weight update in an encrypted cloud environment. In IESNN, probabilistic learning is used to obtain search ranking for searchable index, and probabilistic query is performed based on ciphertext index, which reduces the computational complexity of query significantly. Compared to traditional SE, it is proposed that adversarial learning and automatic weight update in response to user's timely query of the latest data set without expensive communication overhead. The proposed IESNN performs better than the previous works, bringing the query complexity closer to $O(\log N)$ and introducing low overhead on computation and communication.

\end{abstract}
\section{Introduction}\label{section1}
The frequent and massive disclosure of private data has drawn the growing attention of the public to the \emph{cyberspace security}. Meanwhile, \emph{cloud storage} services are increasingly attracting individuals and enterprises to outsource data into cloud server with the rapid development of \emph{cloud computing}. Unfortunately, outsourcing data into cloud server may reveal the privacy of data~\cite{proc/Wang/2010,proc/Yu/2010}. In \emph{cloud security}, \emph{searchable encryption} (SE) has received widespread attention as it protects the privacy of outsourced data and prevents sensitive information from leaking~\cite{article/Kumar/2019}. However, traditional SE~\cite{article/Cao/2014,article/Guo/2018,article/Li/2014,proc/Song/2000,article/Sun/2014,proc/Wang/2010,proc/Wong/2009,article/Xia/2016,proc/Yu/2010} requires high computation and communication overhead to enable \emph{dynamic search} and \emph{dynamic update}, which makes SE still unable to satisfy user's experience and requirements of the actual application adequately. Actually, \emph{machine learning} (ML) can provide intelligent and efficient means yet the current popular ML only supports plaintext data training and can not satisfy the special requirements of encrypted cloud data. Therefore, it is necessary to discuss the cross-fusion problem of ML and SE, and introduce intelligence and high-efficiency into SE.

SE has been continuously developed since it was proposed~\cite{proc/Song/2000}, and \emph{multi-keyword ranked search} scheme is recognized as excellent~\cite{article/Kumar/2019}. Cao et al.~\cite{article/Cao/2014} first discussed privacy-preserving multi-keyword ranked search over encrypted cloud data (MRSE) for single data owner model, and established strict privacy requirements. They first used \emph{asymmetric scalar-product preserving encryption} (ASPE)~\cite{proc/Wong/2009} to obtain the \emph{similarity score} of the query vector and the index vector. In this way, cloud server can retrieve \emph{top-k} documents that are most relevant to the data user's query request. However, since matrix operations require high computation overhead, MRSE is not suitable for practical application scenario. For the purpose of managing the keyword dictionary dynamically and improving system performance, Li et al.~\cite{article/Li/2014} proposed efficient multi-keyword ranked query over encrypted data in cloud computing (MKQE) based on MRSE, which owns a low overhead index construction algorithm and a novel trapdoor generation algorithm. However, it still has no major breakthrough in improving search efficiency when the data set is large. To achieve \emph{dynamic search}, Xia et al.~\cite{article/Xia/2016} provided a secure and dynamic multi-keyword ranked search scheme over encrypted cloud data (EDMRS) to support dynamic operation in SE. For tree-based index structures, search efficiency is improved by the greedy depth-first search (GDFS) algorithm and parallel computing. Regrettably, the search efficiency of ordinary balanced binary tree they used gradually decreases and tends to linear search efficiency when migrating to multiple data owners model with large amount of differential data. Moreover, maintaining such an index tree is not flexible and efficient. Guo et al.~\cite{article/Guo/2018} discussed secure multi-keyword ranked search over encrypted cloud data for multiple data owners model (MKRS\_MO) and designed a heuristic weight generation algorithm based on the relationships among keywords, documents and owners (KDO). They considered the correlation among documents and the impact of documents' quality on search results. Experiments on the real-world data set showed that MKRS\_MO is better than the schemes using traditional $TF\times IDF$ keyword weight model~\cite{article/Sun/2014}. However, the fly in the ointment is that the operations of calculating index similarity in MKRS\_MO may lead to ``curse of dimensionality", which limits the availability of the system. Last but not least, they ignored the secure solution in \textit{known background model}~\cite{article/Cao/2014} (\emph{threat model} for measuring the ability of cloud server to evaluate private data and the risk of revealing private information in SE system).

For the first time, this paper proposes \emph{interpretable encrypted searchable neural networks} (IESNN) to explore \emph{intelligent SE}. Based on the \emph{neural network}, we propose \emph{sorting network} and employ \emph{probabilistic learning} to obtain the query ranking for encrypted searchable index. To be specific, firstly it performs a sufficient amount of random queries (obey \emph{uniform distribution}) and then calculates the sum of the inner product of each index vector and all random query vectors. Finally it sorts the index vectors according to the match scores from high to low. Therefore, the probabilistic ranking of the index is close to the ranking in the actual query, which reduces the computational complexity of the query significantly. Moreover, \emph{probabilistic query} with computational complexity close to $O(\log N)$, is used to retrieve \emph{top-k} documents. In order to achieve secure weight update without revealing private information to ``semi-trusted" cloud server~\cite{proc/Wang/2010,proc/Yu/2010}, we propose \emph{searching adversarial network} and \emph{weight update network} in an encrypted cloud environment. Specifically, in order to respond to user's timely query of the latest data set, we employ \emph{adversarial learning}~\cite{article/Goodfellow/2014} and \emph{optimal game equilibrium} to make the probabilistic ranking of the index close to its popular ranking. Furthermore, we combine \emph{backpropagation neural network}~\cite{article/Hinton/2006} with \emph{discrete Hopfield neural network}~\cite{article/Park/1993} to enable \emph{automatic weight update}. It is worth mentioning that the update operations are done in the cloud, which means there is no expensive communication overhead. So we can use IESNN for model training and intelligent system implementation. On the one hand, it introduces intelligence into the SE system, which improves user's experience and reduces system overhead. On the other hand, training data sources for ML can be derived from ciphertext. It means that data mining based on ciphertext analysis can not only obtain results consistent with plaintext analysis but also strengthen the intensity of data privacy protection.
\begin{table}[ht]
\centering
\scriptsize
\caption{Comparison of related works.}\label{tab1}
\begin{tabular}{|c|c|c|c|c|c|}
\hline
Item& MRSE~\cite{article/Cao/2014} &MKRS\_MO~\cite{article/Guo/2018} &MKQE~\cite{article/Li/2014}&EDMRS~\cite{article/Xia/2016}& IESNN\\
\hline
high-precision query&            $\surd$&  $\surd$&  $\surd$&  $\surd$&   $\surd$ \\
privacy-preserving query&        $\surd$&  $\times$& $\surd$&  $\surd$&   $\surd$ \\
automatic weight update&         $\times$& $\times$& $\times$& $\times$&  $\surd$ \\
high-quality ranked search&      $\times$& $\surd$&  $\surd$&  $\times$&  $\surd$ \\
efficient multi-keyword search&  $\surd$&  $\surd$&  $\surd$&  $\surd$&   $\surd$ \\
flexible dynamic maintenance&    $\times$& $\times$& $\times$& $\times$&  $\surd$ \\
\hline
\end{tabular}
\end{table}

\noindent \textbf{\emph{Our main contributions}} are summarized as follows:
\begin{description}
  \item[(1)] Towards \emph{intelligent SE} by combining popular ML with traditional SE effectively;
  \item[(2)] We employ \emph{probabilistic learning} method to achieve \emph{maximum likelihood searching} and improve search efficiency significantly;
  \item[(3)] We use IESNN to implement \emph{flexible dynamic operation and maintenance} in an encrypted cloud environment.
\end{description}

The remainder of this paper is organized as follows: Section~\ref{Section2} describes the SE model. Section~\ref{Section3} describes the details of IESNN and its performance tests. Section~\ref{Section4} discusses our solution and its implications.
\section{Searchable Encryption Model}\label{Section2}
\subsection{System Model}
\begin{figure}[ht]
\scriptsize
\centering
\includegraphics[width=10.5cm]{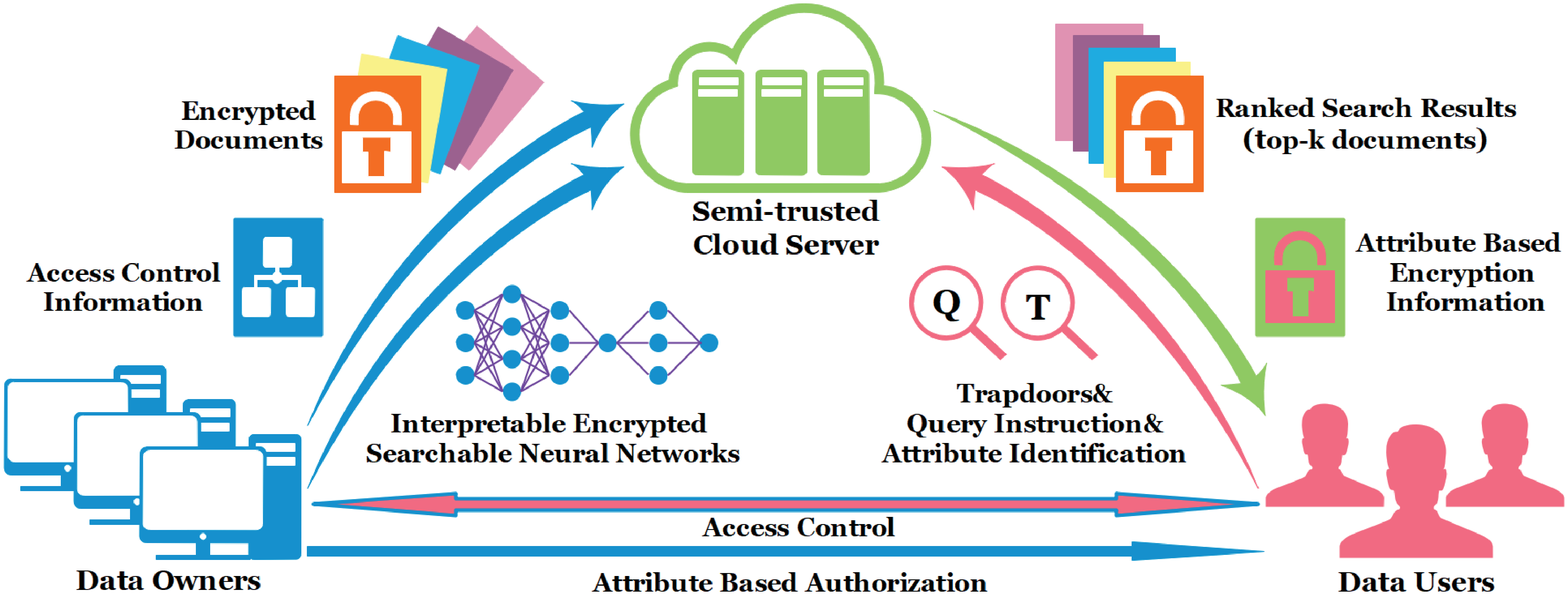}\\
\caption{The basic architecture of searchable encryption system}\label{fig:1}
\end{figure}

The system model proposed in this paper consists of three parties, is depicted in Fig.~\ref{fig:1}, and the specific description is as follows:
\begin{description}
  \item[Data owners($DO$):] $DO$ are responsible for building searchable index and original IESNN, encrypting the data and sending them to cloud server.
  \item[Data users($DU$):] $DU$ are consumers of cloud services. Once the license is granted, they can retrieve the encrypted cloud data.
  \item[Cloud server($CS$):] $CS$ is considered``semi-trusted"in SE~\cite{proc/Wang/2010,proc/Yu/2010}. It provides cloud service, including running authorized access controls, performing searches for encrypted cloud data based on query requests, returning \emph{top-k} documents to $DU$ and enabling dynamic operation and maintenance with IESNN.
\end{description}
\subsection{System Framework}
\begin{description}
  \item[\textit{Setup}:] Based on privacy requirements in \textit{known background model}~\cite{article/Cao/2014}, $DO_{i}$ determines the size $N_{i}$ of dictionary $D_{i}$, the number $U_{i}$ of pseudo-keyword, sets the parameter $V_{i} = U_{i} + N_{i}$. For all data owners $DO$ = \{$DO_{1}$,\ldots,$DO_{m}$\}, we have $V$ = \{$V_{1}$,\ldots,$V_{m}$\}, $U$ = \{$U_{1}$,\ldots,$U_{m}$\}, $N$ = \{$N_{1}$,\ldots,$N_{m}$\}.

  \item[\textit{KeyGen}($V$):] $DO$ generate secret key $SK$ = \{$SK_{1}$,\ldots,$SK_{s}$\},
  where $SK_{i}$ = \{$S_{i}$, $M_{i,1}$, $M_{i,2}$\}, $M_{i,1}$ and $M_{i,2}$ are two invertible matrices with the dimension $V_{i} \times V_{i}$  and $S_{i}$ is a random $V_{i}$-length vector.

  \item[\textit{Extended-KeyGen}($SK_{i},Z_{i}$):] For \emph{dynamic search}~\cite{article/Li/2014}, if $Z_{i}$ new keywords are added into the $i$-th dictionary $D_{i}$,  $DO_{i}$ generates a new $SK_{i}'$ = \{$S_{i}'$, $M_{i,1}'$, $M_{i,2}'$\}, two invertible matrices $M_{i,1}'$ and $M_{i,2}'$ with the dimension $(V_{i}+Z_{i}) \times (V_{i}+Z_{i})$, and a new $(V_{i} + Z_{i})$-length vector $S_{i}'$.

  \item[\textit{BuildIndex}($I,SK$):] In order to reduce the possibility that ``semi-trusted" cloud server~\cite{proc/Wang/2010,proc/Yu/2010} evaluates the private data successfully, $DO$ first build searchable indexes for documents and obtain the weighted index vectors, and then fill index vectors with random pseudo-keywords (obey \emph{Gaussian distribution}) and obtain secure index vectors with high privacy protection strength~\cite{article/Cao/2014}. Finally they use secure index vectors to build IESNN ($\mathcal{I}$) and send $\mathcal{I}$ to $CS$ (specific example: $DO_{i}$ ``splits" index vector $I_{i}$ into two random vectors $\{I_{i,1},I_{i,2}\}$. Specifically, if $S_{i}[t]=0$, $I_{i,1}[t]$ = $I_{i,2}[t]$ = $I_{i}[t]$ ; else if $S_{i}[t]=1$, $I_{i,1}[t]$ is a random value, $I_{i,2}[t] = I_{i}[t]-I_{i,1}[t]$. $DO_{i}$ encrypts $I_{i}$ as $\widetilde{I}_{i}$ = $\{ M_{i,1}^T I_{i,1}, M_{i,2}^T I_{i,2} \} $ with $SK_{i}$).

  \item[\textit{Trapdoor}($Q,SK$):] $DU$ send query request (query keywords and $k$) to $DO$. $DO$ generate query $Q$ = \{$Q_{1}$,\ldots,$Q_{m}$\} (where $Q_{i}$ is a weighted vector with dimension $V_{i}$) and calculate the trapdoor $T$ = \{$T_{1}$,\ldots,$T_{m}$\} using $SK$ and send $T$ to $DU$ (specific example: $DO_{i}$ ``splits" query vector $Q_{i}$ into two random vectors $\{Q_{i,1},Q_{i,2}\}$. Specifically, if $S_{i}[t]=0$, $Q_{i,1}[t]$ is a random value, and $Q_{i,2}[t]=Q_{i}[t]-Q_{i,1}[t]$; else if $S_{i}[t]=1$, $Q_{i,1}[t]=Q_{i,2}[t]=Q_{i}[t]$. Finally, $DO_{i}$ encrypts $Q_{i}$ as $T_{i}$ = $\{M^{-1}_{i,1} Q_{i,1}, M^{-1}_{i,2} Q_{i,2}\}$ with $SK_{i}$).

  \item[\textit{Query}($\mathcal{I},T,k)$:]  $DU$ send trapdoors, query instruction and attribute identification to $CS$. $CS$ performs searches based on the query, and returns \emph{top-k} documents to $DU$.
\end{description}
\section{Interpretable Encrypted Searchable Neural Networks}\label{Section3}
\subsection{Maximum Likelihood Searching}
\begin{figure}[ht]
\small
\centering
\includegraphics[width=11cm]{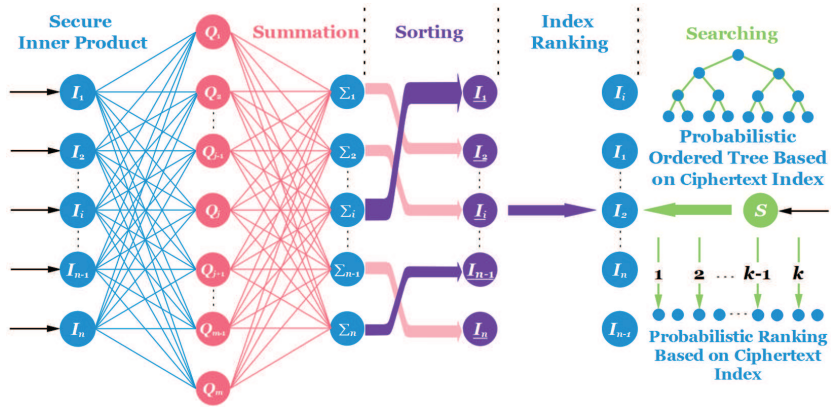}\\
\caption{Sorting network}\label{fig:2}
\end{figure}

We employ \emph{inner product similarity}~\cite{book/Witten/1999} to quantitatively evaluate the effective similarity between the query vector and the index vector. As illustrated in Fig.~\ref{fig:2} (for an intuitive understanding, it shows the unencrypted network), in \emph{sorting network}, it performs a sufficient amount of random queries (obey \emph{uniform distribution}: $X \sim U(-\sigma\sqrt{3},\sigma\sqrt{3})$, that is $f(x) = \frac{1}{2\sigma\sqrt{3}},x \in [-\sigma\sqrt{3},\sigma\sqrt{3}]$), and then calculates the sum of the inner product $\sum^m_{j=1} I^{T}_{i}\cdot Q_{j}$ of each index vector and all random query vectors with formula~\ref{eq:1}. Finally it sorts the index vectors according to the match scores from high to low. Therefore, the index ranking obtained by \emph{probabilistic learning} is close to the ranking in the actual query.
\begin{equation}\label{eq:1}
Score = \widetilde{I}_{i}\cdot T_{i} = \{ M_{i,1}^{T} I_{i,1}, M_{i,2}^{T} I_{i,2} \} \cdot \{ M_{i,1}^{-1}Q_{i,1}, M_{i,2}^{-1}Q_{i,2} \} = I^{T}_{i}\cdot Q_{i}
\end{equation}

\begin{figure}[ht]
\scriptsize
\centering
\includegraphics[width=6cm]{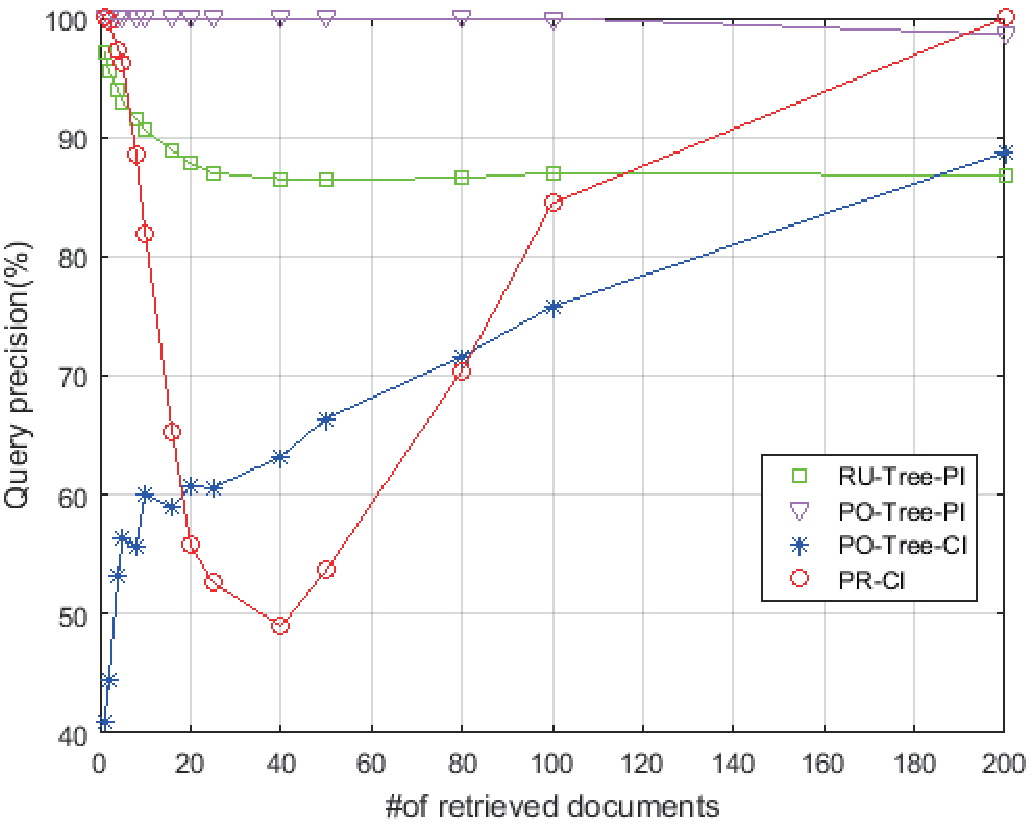}
\includegraphics[width=6cm]{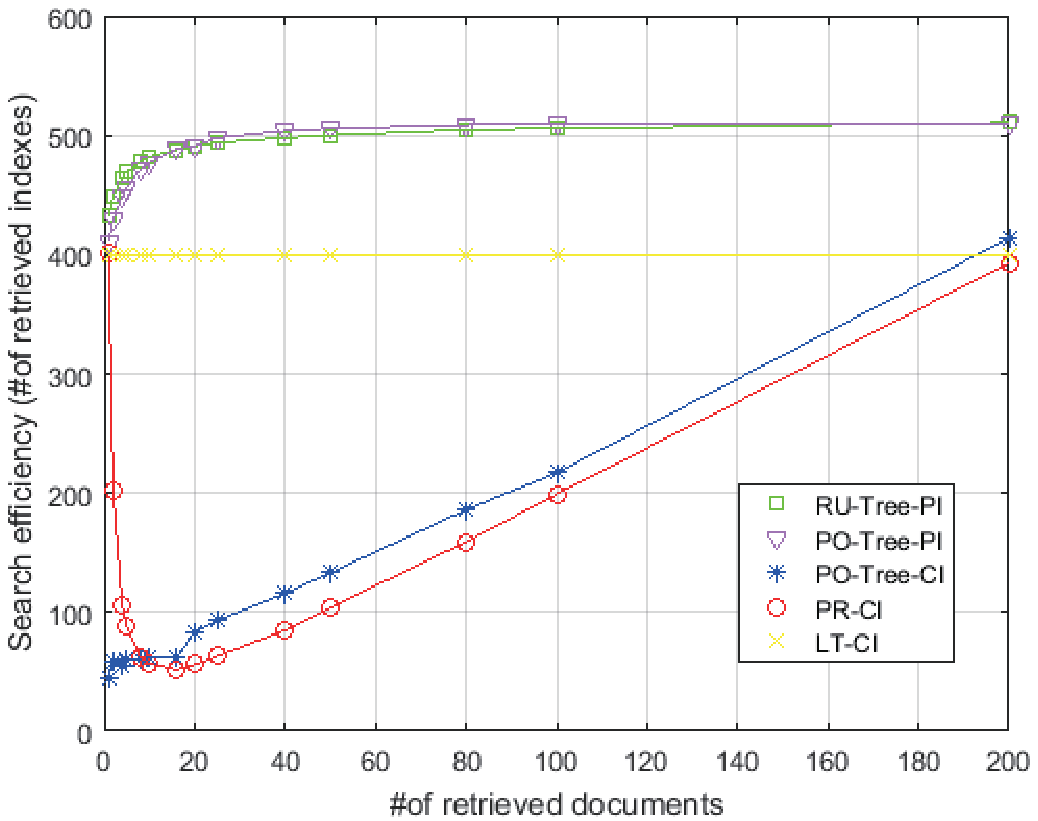}\\
\caption{\scriptsize \textbf{Performance testing of multiple query algorithms}. \{The \emph{query precision} and \emph{search efficiency} for different numbers of retrieved documents with the same document collection (400) and dictionary (2,000). It requires an average of 100 experimental results to measure performance of the following subjects: random unordered tree based on plaintext index(RU-Tree-PI)~\cite{article/Guo/2018,article/Xia/2016}, probabilistic ordered tree based on plaintext index(PO-Tree-PI), probabilistic ordered tree based on ciphertext index(PO-Tree-CI), probabilistic ranking based on ciphertext index(PR-CI), linear traversal based on ciphertext index(LT-CI)~\cite{article/Cao/2014,article/Li/2014}. The number of retrieved \emph{top-k} documents is the factor of 400: $k = 1,2,4,5,8,10,16,20,25,40,50,80,100,200.$ Note: the nodes of the tree are also included in the number of retrieval indexes. According to the experimental results, \emph{probabilistic query} can significantly improve the search efficiency. When $k$ takes a specific interval value, the query precision is high or low. It is because that the probabilistic ranking of the index vector is not strictly ordered, and the query is random. Therefore, when the query vector is very ``unpopular", the query precision will become lower, and when the query vector is ``popular", the query precision and search efficiency will perform well.\}}\label{fig:3}
\end{figure}

\noindent We implement the proposed scheme using Python in Windows 10 operation system with Intel Core i5 Processor 2.40GHz and test its efficiency on a real-world document set (IEEE INFOCOM publications, including 400 papers and 2,000 keywords). The \emph{probabilistic query} algorithm based on the probabilistic ranking of encrypted searchable index brings the query complexity closer to $O(\log N)$. As shown in Fig.~\ref{fig:3}, when retrieving the same number of \emph{top-k} documents, \emph{probabilistic query} performs better than the related works that based on tree search~\cite{article/Guo/2018,article/Xia/2016} and matrix operation~\cite{article/Cao/2014,article/Li/2014}. As the ordered feature of the balanced binary tree is not guaranteed in the index tree and the query based on matrix operation needs to traverse all indexes to retrieve \emph{top-k} documents, the number of retrieved indexes is far more than the number of retrieved documents.
\subsection{Adversarial Learning}
\begin{figure}[ht]
\small
\centering
\includegraphics[width=11cm]{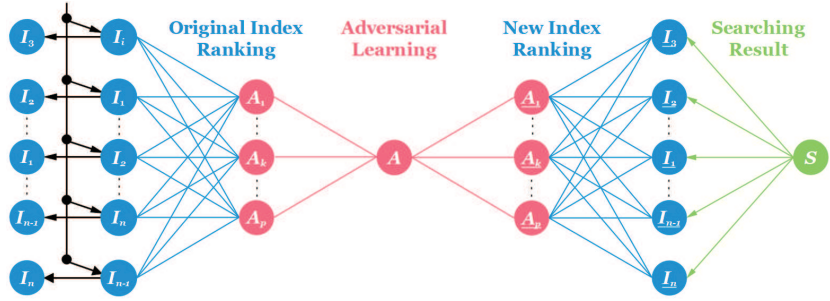}\\
\caption{Searching adversarial networks}\label{fig:4}
\end{figure}

\emph{Adversarial network} works when the probabilistic ranking of the index deviates from the index ranking in the actual query result. As shown in Fig.~\ref{fig:4}, it employs \emph{optimal game equilibrium} to make the probabilistic ranking of the index close to its popular ranking (described by formula~\ref{eq:2}, $p_{i}(x)$ and $p_{q}(y)$ are the \emph{probability distributions} of the index ranking and query result, respectively).
\begin{equation}\label{eq:2}
\min_{S}\max_{A}V(A,S)= \mathbb{E}_{x\sim p_{i}(x)}[\log A(x)] + \mathbb{E}_{y\sim p_{q}(y)}[\log (1 - A(S(y)))]
\end{equation}

\noindent Inspired by \emph{generative adversarial networks} (GAN)~\cite{article/Goodfellow/2014} and \emph{self-attention generative adversarial networks} (SAGAN)~\cite{proc/Zhang/2019} but different from GAN and SAGAN, \emph{searching adversarial networks} (SAN) do not require complex gradient calculations and extensive iterative training. As a matter of fact, SAN only require simple residual calculations and index sorting floating steps. Specifically, after completing the query, the ranked search result list is feedback to \emph{adversarial network} in SAN. \emph{Adversarial network} calculates the residual before and after the weight change of the index corresponding to \emph{top-k} documents, and calculates the relative floating of the index ranking of the feedback result(i.e. new index ranking) and the original index ranking. Finally, SAN send the results of the calculation (the residual of the weights) and the index ranking changes to the \emph{weight update network} as a target for index update (see Fig.~\ref{fig:5} for details).
\subsection{Automatic Weight Update}
\begin{figure}[ht]
\small
\centering
\includegraphics[width=10.5cm]{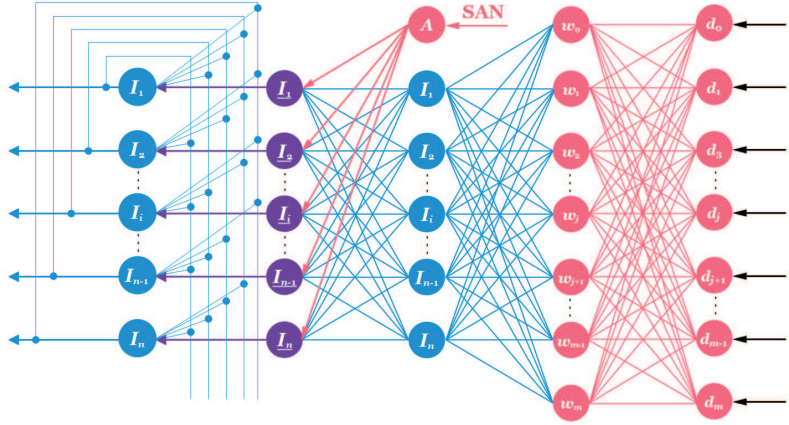}\\
\caption{Weight update network}\label{fig:5}
\end{figure}

As illustrated in Fig.~\ref{fig:5}, in order to achieve \emph{automatic weight update} and respond to users' queries for the latest data sets in a timely manner, \emph{weight update network} (WUN) combines \emph{backpropagation neural network} (BPNN)~\cite{article/Hinton/2006} with \emph{discrete Hopfield neural network} (DHNN)~\cite{article/Park/1993}. In WUN, the update of index weights uses vector and matrix operations to approximate the actual increments, which has the characteristics of local homomorphism for ciphertext operations and plaintext operations. For instance, considering index vector $I_{\alpha} = [\alpha_{1},\ldots,\alpha_{n}]^T,$ query vector $Q_{\gamma} = [\gamma_{1},\ldots,\gamma_{n}]^T,$ and two invertible matrices $M = {(a_{ij})}_{n\times n}$ and $M^{-1} = {(b_{ij})}_{n\times n}.$ The \emph{update principle} of ciphertext index is as follows:

\emph{Matrix and vector multiplication}: $I^T_{\alpha} M = [\sum^n\limits_{i = 1}\alpha_{i}a_{i1},\ldots,\sum^n\limits_{i = 1}\alpha_{i}a_{in}], M^{-1} Q_{\gamma} = [\sum^n\limits_{j = 1}b_{1j}\gamma_{j},\ldots,\sum^n\limits_{j = 1}b_{nj}\gamma_{j}];
$\emph{Secure inner product calculation}: $ I^T_{\alpha} M \cdot M^{-1} Q_{\gamma} = I^T_{\alpha} \cdot Q_{\gamma} = \sum^n\limits_{i = 1}\alpha_{i}\gamma_{i};$
\emph{Index vector update}: $(I^T_{\alpha} + \Delta I^T_{\alpha}) M = [ \sum^n\limits_{i = 1}(\alpha_{i}+\Delta \alpha_{i})a_{i1},\ldots,\sum^n\limits_{i = 1}(\alpha_{i}+\Delta \alpha_{i})a_{in}] \approx [\sum^n\limits_{i = 1}\alpha_{i}a_{i1} + \Delta \beta_{1},\ldots,\sum^n\limits_{i = 1}\alpha_{i}a_{in}+\Delta \beta_{n}] = I^T_{\alpha}M + \Delta I^T_{\beta}M;
$\emph{Inner product approximation}: $(I^T_{\alpha} + \Delta I^T_{\alpha})M \cdot M^{-1} Q_{\gamma} \approx (I^T_{\alpha}M + \Delta I^T_{\beta}M) \cdot M^{-1} Q_{\gamma} = I^T_{\alpha} \cdot Q_{\gamma} + \Delta I^T_{\beta} \cdot Q_{\gamma} = \sum^n\limits_{i = 1}\alpha_{i}\gamma_{i} + \sum^n\limits_{i = 1}\Delta\beta_{i}\gamma_{i}.$

\noindent \textbf{\textit{Asynchronous work mode of WUN}:} The update task from SAN to WUN is only updating the weight of an index, while other indexes still retain their original weight. i.e.
\begin{equation}\label{eq:5}
I_{j}(t+1) =  \left\{ \begin{array}{ll}
sgn[net_{j}(t)],& j=i\\
I_{j}(t),& j\neq i
\end{array} \right.,\\
I_{j}(t+1) =  \left\{ \begin{array}{ll}
satlins[net_{j}(t)],& j=i\\
I_{j}(t),&j \neq i
\end{array} \right.
\end{equation}

\noindent \textbf{\textit{Synchronous work mode of WUN}:} The synchronous work mode is parallel, i.e. the weights of all indexes are all changed in one update. The adjustment of the weight is determined according to the current input value. The weight update is complete and the weight of an index continues to be used for the next update. When the weight of each index is stabilized, the work ends.
\begin{equation}\label{eq:6}
\left\{ \begin{array}{ll}
I_{j}(t+1) =  sgn[net_{j}(t)],& j=1,2,\ldots,n\\
I_{j}(t+1) =  satlins[net_{j}(t)],&j=1,2,\ldots,n
\end{array} \right.
\end{equation}

\noindent When updating an index, the schemes~\cite{article/Guo/2018,article/Xia/2016} employ tree-based index need to update the index vector itself (leaf node of index tree) and its corresponding other data (parent node of leaf node). Moreover, in order to achieve \emph{dynamic search}, the current schemes~\cite{article/Cao/2014,article/Guo/2018,article/Li/2014,article/Sun/2014,article/Xia/2016} need to download the ciphertext index from the cloud, update its plaintext after local decryption, and finally upload the new ciphertext index to the cloud. In comparison, our solution only needs to update the index vector in the cloud with touching a smaller amount of data and introduce low overhead on computation and communication.
\subsection{Overall Operation and Maintenance of IESNN}
\begin{figure}[ht]
\small
\centering
\includegraphics[width=10cm]{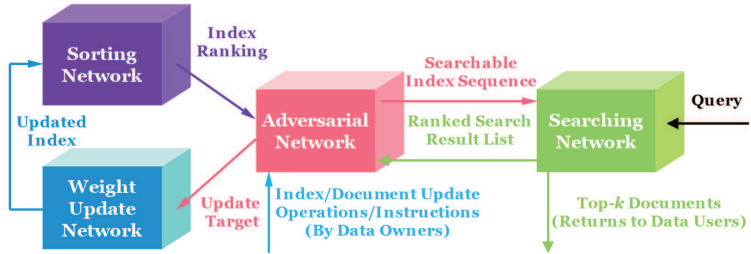}\\
\caption{The overall composition of IESNN}\label{fig:6}
\end{figure}

As shown in Fig.~\ref{fig:6}, IESNN consist of \emph{sorting net}, \emph{adversarial net}, \emph{searching net} and \emph{weight update net}. Except that the initial index weight needs to be generated by data owners, the rest of automatic update operations (``add, delete, change and investigate" operations of index) are all completed in an encrypted cloud environment. The system forms a ``query-learning-update-learning-query" \emph{self-attention}~\cite{proc/Zhang/2019} loop and an ``automatic operation and maintenance" mechanism. \emph{Dynamic operation and maintenance} of SE system are almost entirely done in cloud server. On the one hand, implementing \emph{dynamic operation and maintenance} in an encrypted cloud environment not only improves the usability and flexibility of SE system, but also enhances the strength of privacy protection. On the other hand, when it is necessary to update the index in cloud server, compared with traditional SE~\cite{article/Cao/2014,article/Guo/2018,article/Li/2014,article/Sun/2014,article/Xia/2016}, our solution eliminates the need to rebuild the index locally and upload a new index to cover the old index stored in the cloud, which introduces low overhead on computation, communication and local storage.
\section{Discussion}\label{Section4}
In this paper, we discuss the cross-fusion problem of ML and SE, and propose IESNN. We creatively combine popular ML with traditional SE, which is committed to exploring \emph{intelligent SE}. We employ \emph{probabilistic learning} method to generate \emph{sorting network} that is trained by a sufficient amount of random queries, which makes a contribution to achieve \emph{maximum likelihood searching} and bring the query complexity closer to $O(\log N)$. It means that exploiting ML to optimize the query is effective in an \emph{uncertain system}, even better than special construction methods. Obviously, traditional query algorithms based on matrix operations and tree searching are not optimistic in big data environments because high dimensional data processing can lead to ``curse of dimensionality" and even system crashes. Implementing \emph{flexible dynamic operation and maintenance} in an encrypted cloud environment with IESNN that reduces communication overhead, protects data privacy and leverages cloud computing well.
\section*{Acknowledgment}
This work was supported by ``the Fundamental Research Funds for the Central Universities" (No. 30918012204) and ``the National Undergraduate Training Program for Innovation and Entrepreneurship" (Item number: 201810288061). NJUST graduate Scientific Research Training of `Hundred, Thousand and Ten Thousand' Project ``\emph{Research on Intelligent Searchable Encryption Technology}".
\bibliographystyle{splncs04}
\bibliography{mybibliography}

\begin{thebibliography}{10}
\providecommand{\url}[1]{\texttt{#1}}
\providecommand{\urlprefix}{URL }
\providecommand{\doi}[1]{https://doi.org/#1}

\bibitem{article/Cao/2014}
{Cao}, N., {Wang}, C., {Li}, M., {Ren}, K., {Lou}, W.: Privacy-preserving
  multi-keyword ranked search over encrypted cloud data. {IEEE} Trans. Parallel
  Distrib. Syst.  \textbf{25}(1),  222--233 (2014)

\bibitem{article/Goodfellow/2014}
Goodfellow, I.J., Pouget{-}Abadie, J., Mirza, M., Xu, B., Warde{-}Farley, D.,
  Ozair, S., Courville, A.C., Bengio, Y.: Generative adversarial networks. CoRR
   \textbf{abs/1406.2661} (2014)

\bibitem{article/Guo/2018}
Guo, Z., Zhang, H., Sun, C., Wen, Q., Li, W.: Secure multi-keyword ranked
  search over encrypted cloud data for multiple data owners. Journal of Systems
  and Software  \textbf{137}(3),  380--395 (2018)

\bibitem{article/Hinton/2006}
Hinton, G.E., Osindero, S., Welling, M., Teh, Y.W.: Unsupervised discovery of
  nonlinear structure using contrastive backpropagation. Cognitive Science
  \textbf{30}(4),  725--731 (2006)

\bibitem{article/Kumar/2019}
Kumar, D.V.N.S., Thilagam, P.S.: Approaches and challenges of privacy
  preserving search over encrypted data. Inf. Syst.  \textbf{81},  63--81
  (2019)

\bibitem{article/Li/2014}
Li, R., Xu, Z., Kang, W., Yow, K., Xu, C.: Efficient multi-keyword ranked query
  over encrypted data in cloud computing. Future Generation Comp. Syst.
  \textbf{30}(1),  179--190 (2014)

\bibitem{article/Park/1993}
{Park}, J.H., {Kim}, Y.S., {Eom}, I.K., {Lee}, K.Y.: Economic load dispatch for
  piecewise quadratic cost function using hopfield neural network. {IEEE}
  Trans. Power Syst.  \textbf{8}(3),  1030--1038 (1993)

\bibitem{proc/Song/2000}
Song, D.X., Wagner, D.A., Perrig, A.: Practical techniques for searches on
  encrypted data. In: {IEEE} S \& P 2000. pp. 44--55. {IEEE} Computer Society
  (2000)

\bibitem{article/Sun/2014}
{Sun}, W., {Wang}, B., {Cao}, N., {Li}, M., {Lou}, W., {Hou}, Y.T., {Li}, H.:
  Verifiable privacy-preserving multi-keyword text search in the cloud
  supporting similarity-based ranking. {IEEE} Trans. Parallel Distrib. Syst.
  \textbf{25}(11),  3025--3035 (2014)

\bibitem{proc/Wang/2010}
Wang, C., Wang, Q., Ren, K., Lou, W.: Privacy-preserving public auditing for
  data storage security in cloud computing. In: {INFOCOM} 2010. pp. 525--533
  (2010)

\bibitem{book/Witten/1999}
Witten, I.H., Moffat, A., Bell, T.C.: Managing Gigabytes: Compressing and
  Indexing Documents and Images, Second Edition. Morgan Kaufmann (1999)

\bibitem{proc/Wong/2009}
Wong, W.K., Cheung, D.W., Kao, B., Mamoulis, N.: Secure knn computation on
  encrypted databases. In: {ACM} {SIGMOD} 2009. pp. 139--152. {ACM} (2009)

\bibitem{article/Xia/2016}
{Xia}, Z., {Wang}, X., {Sun}, X., {Wang}, Q.: A secure and dynamic
  multi-keyword ranked search scheme over encrypted cloud data. {IEEE} Trans.
  Parallel Distrib. Syst.  \textbf{27}(2),  340--352 (2016)

\bibitem{proc/Yu/2010}
Yu, S., Wang, C., Ren, K., Lou, W.: Achieving secure, scalable, and
  fine-grained data access control in cloud computing. In: {INFOCOM} 2010. pp.
  534--542. {IEEE} (2010)

\bibitem{proc/Zhang/2019}
Zhang, H., Goodfellow, I.J., Metaxas, D.N., Odena, A.: Self-attention
  generative adversarial networks. In: {ICML} 2019. pp. 7354--7363. {PMLR}
  (2019)

\end{thebibliography}
\end{document}